\documentclass[letterpaper, 10 pt, conference]{ieeeconf}  

\usepackage[T1]{fontenc}
\usepackage[utf8]{inputenc}

\renewcommand\labelenumi{(\roman{enumi})}
\renewcommand\theenumi\labelenumi

\usepackage{amsmath}
\usepackage{amsfonts}
\usepackage{nicefrac}
\usepackage[exponent-product=\cdot, per-mode=symbol]{siunitx}
\usepackage{enumerate}

\usepackage{booktabs}
\usepackage{graphicx}
\usepackage{epstopdf}
\usepackage{xcolor}
\definecolor{vgRed}{RGB}{193, 48, 24}
\definecolor{vgOrange}{RGB}{243, 111, 19}
\definecolor{vgYellow}{RGB}{235, 203, 56}
\definecolor{vgGreen}{RGB}{162, 185, 105}
\definecolor{vgLightBlue}{RGB}{13, 149, 188}
\definecolor{vgDarkBlue}{RGB}{6, 56, 81}

\newcommand{\N}{\mathbb{N}}
\newcommand{\C}{\mathbb{C}}

\newcommand{\R}{\mathbb{R}}
\newcommand{\Rp}{\R^+}

\newcommand{\diag}{\operatorname{diag}}
\newcommand{\argmax}{\operatorname{argmax}}

\newcommand{\hinf}{\mathcal{H}_\infty}

\usepackage{acronym}
\acrodef{bem}[BEM]{blade element momentum theory}
\acrodef{lq}[LQ]{linear quadratic}
\acrodef{mpc}[MPC]{Model Predictive Controller}
\acrodef{wt}[WT]{Wind Turbine}
\acrodef{pi}[PI]{proportional-integral}
\acrodef{siso}[SISO]{single-input single-output}
\acrodef{lmi}[LMI]{linear matrix inequality}
\acrodefplural{lmi}[LMIs]{linear matrix inequalities}
\acrodef{del}[DEL]{damage equivalent load}
\acrodef{rms}[RMS]{root-mean-square}
\acrodef{ti}[TI]{turbulence intensity}
\acrodef{tfm}[TFM]{transfer function matrix}
\acrodefplural{tfm}{transfer function matrices}

\usepackage{tikz}
\usepackage{pgfplots}
\usepackage{pgfplotstable}
\pgfplotsset{compat=newest}
\usetikzlibrary{plotmarks}
\usetikzlibrary{fit}
\usetikzlibrary{positioning}
\usetikzlibrary{shapes}
\usetikzlibrary{arrows,automata,plotmarks}
\usetikzlibrary{decorations.pathreplacing,decorations.markings}
\usetikzlibrary{backgrounds}
\usetikzlibrary{calc}
\usetikzlibrary{patterns}
\usetikzlibrary{matrix}
\usepgfplotslibrary{statistics}
\usepgfplotslibrary{groupplots}
\usepgfplotslibrary{external}
\tikzset{external/system call={pdflatex \tikzexternalcheckshellescape -halt-on-error
    -interaction=batchmode -jobname "\image" "\texsource"}}
\newtheorem{problem}{Problem}

\usepackage{todonotes}
\renewcommand{\todo}[2][]{\tikzexternaldisable\@todo[#1]{#2}\tikzexternalenable}

\pgfplotsset{every axis/.append style={semithick,tick style={major tick
            length=4pt,semithick,black}}}

\pgfkeys{/pgfplots/x axis shift down/.style={
        x axis line style={yshift=-#1},
        xtick style={yshift=-#1},
        tick align=outside,
        xticklabel shift={#1}}}

\pgfkeys{/pgfplots/y axis shift left/.style={
        y axis line style={xshift=-#1},
        ytick style={xshift=-#1},
        yticklabel shift={#1},
        scaled y ticks = false,
        tick align=outside,
        y tick label style={/pgf/number format/fixed,
        }
    }
}

\pgfplotsset{myPlot/.style={%
        clip = false,
        grid=both,
        grid style={draw=black!10},
        major tick length=0pt,
        minor tick length=0pt,
        axis lines = box,
        axis line style= {-, draw=black!10, draw opacity=1, line width=0.4pt},
        width=8cm,
        height=4cm,
        line width = 0.7pt,
    }
}
\usepackage{mathtools,stackengine,scalerel}
\newcommand{\rawsat}[3]{\ThisStyle{\raisebox{#1\LMpt}{\kern.5\LMpt\scaleto{\rawsatimg{#3}}{#2\LMex}\kern.5\LMpt}}}
\newcommand{\rawsatimg}[1]{%
\tikzexternaldisable
\begin{tikzpicture}
\coordinate (A) at (-7,-7);
\coordinate (B) at (-2,-7);
\coordinate (C) at (2,7);
\coordinate (D) at (7,7);
\draw [black, line width=#1pt] (A)--(B)--(C)--(D);
\end{tikzpicture}%
\tikzexternalenable
}
\makeatletter
\newcommand\sat{
    \relax\if@display
        \mathop{\rawsat{-5}{3.2}{20}}
    \else
        \mathop{\rawsat{-2.4}{2.25}{30}}
    \fi
}

\pgfdeclareshape{satnode}{
\inheritsavedanchors[from={rectangle}]
\inheritbackgroundpath[from={rectangle}]
\inheritanchorborder[from={rectangle}]
\foreach \x in {center,north east,north west,north,south,south east,south west, west, east}{
\inheritanchor[from={rectangle}]{\x}
}
\foregroundpath{
\pgfpointdiff{\northeast}{\southwest}
\pgf@xa=\pgf@x \pgf@ya=\pgf@y
\northeast
\pgfpathmoveto{\pgfpoint{0}{0.45\pgf@ya}}
\pgfpathlineto{\pgfpoint{0}{-0.45\pgf@ya}}
\pgfpathmoveto{\pgfpoint{0.45\pgf@xa}{0}}
\pgfpathlineto{\pgfpoint{-0.45\pgf@xa}{0}}
\pgfpathmoveto{\pgfpointadd{\southwest}{\pgfpoint{-0.2\pgf@xa}{-0.3\pgf@ya}}}
\pgfpathlineto{\pgfpointadd{\southwest}{\pgfpoint{-0.5\pgf@xa}{-0.3\pgf@ya}}}
\pgfpathlineto{\pgfpointadd{\northeast}{\pgfpoint{-0.5\pgf@xa}{-0.3\pgf@ya}}}
\pgfpathlineto{\pgfpointadd{\northeast}{\pgfpoint{-0.4\pgf@xa}{-0.3\pgf@ya}}}
{
   \pgftransformshift{\pgfpointadd{\northeast}{\pgfpoint{-0.4\pgf@xa}{-0.3\pgf@ya}}}
   \pgftransformscale{0.5}
   \pgfsetcolor{black}
}
\pgfusepath{stroke}
}
}

\pgfdeclareshape{lowpassnode}{
\inheritsavedanchors[from={rectangle}]
\inheritbackgroundpath[from={rectangle}]
\inheritanchorborder[from={rectangle}]
\foreach \x in {center,north east,north west,north,south,south east,south west, west, east}{
\inheritanchor[from={rectangle}]{\x}
}
\foregroundpath{
\pgfpointdiff{\northeast}{\southwest}
\pgf@xa=\pgf@x \pgf@ya=\pgf@y
\northeast
\pgfpathmoveto{\pgfpoint{0.4\pgf@xa}{-0.2\pgf@ya}}
\pgfpathlineto{\pgfpoint{0}{-0.2\pgf@ya}}
\pgfpathlineto{\pgfpoint{-0.4\pgf@xa}{0.2\pgf@ya}}
\pgfusepath{stroke}
}
}

\makeatother

\IEEEoverridecommandlockouts                              

\title{\LARGE\bf$\hinf$ Loop-shaping for Power Tracking Control of Wind Turbines*}

\author{Aaron Grapentin$^1$, Christian A. Hans$^2$, and J\"org Raisch$^3$
\thanks{*This work was partially supported by the German Federal Ministry for Economic Affairs and and Climate Action (BMWK), project no. 03EE2036C.}
\thanks{$^{1}$Aaron Grapentin is with the Control Systems Group, Technische Universit\"at Berlin, Germany, {\tt\small grapentin@control.tu-berlin.de}}
\thanks{$^{2}$Christian A. Hans is with the Automation and Sensorics in Networked Systems Group, University of Kassel, Germany, {\tt\small hans@uni-kassel.de}}
\thanks{$^{3}$J\"org Raisch is with the Control Systems Group, Technische Universit\"at Berlin, Germany and Science of Intelligence, Research Cluster of Excellence, Berlin, Germany, {\tt\small raisch@control.tu-berlin.de}}%
\thanks{We thank Arnold Sterle for fruitful discussions and insightful comments.}%
}

\begin{document}

\maketitle
\thispagestyle{empty}
\pagestyle{empty}

\begin{abstract}
In this paper, we present an advanced wind turbine control scheme for power maximization as well as for active power control, which is designed using $\hinf$ loop-shaping.
Our approach involves the synthesis of two separate controllers for two different operating modes.
To ensure smooth transitions between these modes, we implement a bumpless transfer strategy that reduces transient effects.
A comprehensive case study demonstrates the efficacy of our control scheme, showing significant improvements in power tracking accuracy and a reduction in mechanical wear.
Moreover, our control strategy comes with robust stability guarantees.
\end{abstract}
\acresetall


\section{Introduction}
In recent years, wind power has become increasingly important in the overall energy mix.
The growth in wind power generation is driven by technological advances, policy incentives, and a global push towards renewable energy sources to combat climate change \cite{murdock2021renewables}.
As the share of wind energy continues to grow, stricter regulations and operational requirements become necessary.
These include active power tracking and provision of auxiliary services.
Active power-reference tracking involves the ability of a wind farm to dynamically adjust its power output according to power setpoints from grid operators \cite{de2007connection}.
This ensures that the power supplied by wind farms aligns with operational requirements and helps to maintain a balance of generation and demand.
In \cite{gao2016comparison,vrana2018wind} the specific requirements for active power control are analyzed and compared for different markets such as China, the United States, Germany, Denmark, and others.
In this context, accurate power tracking emerges as a critical requirement to enable large shares of wind power generation.
Moreover, auxiliary services offered by wind farms, such as frequency regulation, are gaining prominence and will have significant relevance in practical use cases \cite{vidyanandan2012primary}.
Many of them rely on accurate active power tracking which underlines the importance of this topic.

Various strategies to control wind turbines have been explored.
Wind turbine power maximization has been widely investigated, e.g., in \cite{barambones2010wind,qiao2008wind,vlad2010output}, using different control approaches such as optimal torque control \cite{morimoto2005power}.
In recent years, active power control has gained increasing attention, e.g., in \cite{poschke2020load}, where control strategies are deduced with a focus on reducing mechanical loads and wear.
In \cite{schlipf2016lidar}, active power control performance of state-of-the-art control setups is thoroughly analyzed.
However, aforementioned approaches rely on multiple difficult to tune single-input single-output control loops, resulting in a complex setup.
In \cite{pham2012lqr}, a linear quadratic optimal control approach for wind turbines is proposed.
Relying on wind estimates, the approach comes with a slightly better power tracking accuracy than the baseline controller from \cite{nam2011feedforward}.
Approaches like $\hinf$ loop-shaping have also been investigated in a wind power context, e.g., in \cite{lackner2011structural} to control the platform movement of a floating wind turbine.
Moreover, in \cite{gryning2015wind}, $\hinf$ loop-shaping is utilized for a wind turbine inverter, showcasing its usefulness across various components.
However, despite good performance for other applications, to the author's knowledge, $\hinf$ control has not yet been employed for power maximization and active power control.

In this paper, $\hinf$ control synthesis for wind turbine power maximization and active power control is investigated.
We exploit loop-shaping to introduce integral error states, which enable offset-free tracking of a power reference signal, allowing flexible active power control.
Additionally, we provide robustness guarantees for a set of operating points.
Motivated by \cite{chen2015bumpless}, we derive a bumpless-transfer scheme to smoothly facilitate the transition between power maximization and power reference tracking.
These innovations allows us to greatly enhance power tracking accuracy while reducing mechanical wear on components.
Our scheme offers adaptability to various turbine models and environmental requirements, making it a versatile solution for the evolving landscape of wind energy generators.
In addition, a case study using OpenFAST is contributed delivering accurate simulation results \cite{jonkman2022openfast}.

This paper is structured as follows.
In Section~\ref{sec:model}, the plant model is introduced.
In Section~\ref{sec:controller}, our novel approach for wind turbine controller synthesis using $\hinf$ loop-shaping is presented.
In Section~\ref{sec:transfer}, the bumpless transition between power maximization and power tracking is discussed.
In Section~\ref{sec:case_study}, the controller is evaluated in a case study using OpenFAST \cite{jonkman2022openfast}.
Section~\ref{sec:conclusion} concludes this work.

\subsection{Notation}
The sets of positive integers and nonnegative integers are $\N$ and $\N_0$, respectively.
The set of real numbers is $\R$, and the set of positive real numbers $\Rp$.
The set of complex numbers is $\C$.
The $n\!\times\!n$ identity matrix, with $n\!\in\!\N$, is $I_n$.


\section{Wind Turbine Model}
\label{sec:model}
For controller synthesis, we rely on a standard nonlinear wind turbine model.
While more complex and accurate wind turbine models are available, e.g., OpenFAST, which will be used for the case study \cite{jonkman2022openfast}, a first principles model is used to simplify control synthesis.
This model includes the rotor and generator dynamics, described in Section~\ref{sec:rotor_dynamics}, as well as the fore-aft tower top dynamics, discussed in Section~\ref{sec:tower_dynamics}.
The combined model is described in Section~\ref{sec:nonlinear_model}.
The control inputs are pitch angle $\theta(t)\in\R$ and generator torque $M_g(t)\in\Rp$ at time $t\in\R$, while the wind speed $V(t)\in\Rp$ is an uncertain input.

\subsection{Rotor and Generator Dynamics}
\label{sec:rotor_dynamics}
We assume that rotor and generator shaft are rigidly connected through a gearbox. 
The generator angular speed is $\omega(t) = N_g\omega_r(t)$, with rotor angular speed $\omega_r(t)\in\Rp$ and gearbox ratio $N_g\in\Rp$.
Its dynamics are \cite{burton2011wind}
\begin{equation}
\dot{\omega}(t) = \frac{\rho\pi r^2 N_g^2}{2 J_t}\frac{V(t)^3}{\omega(t)}C_P\big(\lambda(t), \theta(t)\big) - \frac{N_g^2}{J_t}M_g(t),
\label{eq:omega_dynamic}
\end{equation}
with inertia $J_t\in\Rp$, air density $\rho\in\Rp$, rotor area $\pi r^2$ for radius $r\in\Rp$, power coefficient $C_p(\lambda, \theta)\in\Rp$ and tip-speed ratio $\lambda(t) = r\frac{\omega_r(t)}{V(t)}=\frac{r}{N_g}\frac{\omega(t)}{V(t)}$.
Note that $C_p(\cdot)$ is a nonlinear function that depends on blade geometry.

\subsection{Fore-aft Tower Top Dynamics}
\label{sec:tower_dynamics}
The fore-aft movement of the tower top is modelled as a mass-spring-damper system \cite{mirzaei2013}, i.e.,
\begin{equation}
\ddot{x}_t(t) = \frac{\rho\pi r^2}{2M_t}V(t)^2C_T\big(\lambda(t), \theta(t)\big)-\frac{D_t}{M_t}\dot{x}_t(t)-\frac{K_t}{M_t}x_t(t),
\end{equation}
with fore-aft tower top position $x_t(t)\in\R$ and velocity $v_t(t)=\dot{x}_t(t)\in\R$.
The constants $M_t, D_t, K_t\in\Rp$ denote the tower top mass, damping, and stiffness, respectively.
Additionally, the nonlinear thrust coefficient is $C_T(\cdot)\in\Rp$.

\subsection{Nonlinear State Model}
\label{sec:nonlinear_model}
We will use a nonlinear wind turbine model with
\begin{itemize}
    \item states $x(t)=\begin{bmatrix}\omega(t)& x_t(t)& v_t(t)\end{bmatrix}^T,$
    \item control inputs $u(t)=\begin{bmatrix}\theta(t)& M_g(t)\end{bmatrix}^T,$ and
    \item outputs $y(t)=\begin{bmatrix}\omega(t)& \lambda(t)& P(t)& x_t(t)\end{bmatrix}^T,$
\end{itemize}
where $P(t)=\eta\omega(t)M_g(t)$ is the electrical active power, which depends on the generator efficiency $\eta\in\Rp$.
The nonlinear state model reads
\begin{subequations}
\begin{align}
\dot{x}(t)&=f\left(x(t), u(t), V(t)\right)\label{eq:state_equation}\\
&=\begin{bmatrix}
\frac{\rho\pi r^2 N_g^2}{2 J_t}\frac{V(t)^3}{\omega(t)}C_P\big(\lambda(t), \theta(t)\big) - \frac{N_g^2}{J_t}M_g(t)\\
v_t(t)\\
\frac{\rho\pi r^2}{2M_t}V(t)^2C_T\big(\lambda(t), \theta(t)\big)-\frac{D_t}{M_t}v_t(t)-\frac{K_t}{M_t}x_t(t)
\end{bmatrix},\nonumber\\
y(t)&=g\left(x(t), u(t), V(t)\right)\\
&=\begin{bmatrix}\omega(t)&\frac{r}{N_g}\frac{\omega(t)}{V(t)}&\eta\omega(t)M_g(t)&x_t(t)\end{bmatrix}^T.\nonumber
\label{eq:output_equation}
\end{align}
\label{eq:nonlinear_model}
\end{subequations}


\section{Controller Synthesis}
\label{sec:controller}
\begin{figure}[htbp]
	\centering
	\includegraphics{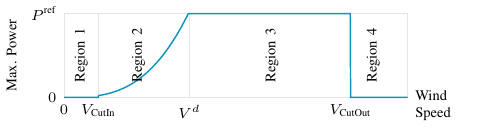}
	\caption{Operating regions of wind turbine.}
	\label{fig:regionData}
\end{figure}

A wind turbine exhibits 4 characteristic operating regions as shown in Figure~\ref{fig:regionData}
In regions~2 and 3, it is enabled pursuing different objectives.
Therefore often two different controllers are used: one for power maximization in region~2 and one for power tracking in region~3.
We follow the same approach in our work, but will additionally ensure bumpless transfer between the two controllers (Section~\ref{sec:transfer}).
In what follows, we perform control synthesis using the loop-shaping approach described in \cite{skogestad2005multivariable} as it allows to handle the impact of disturbances and linearization errors.
In Section~\ref{sec:equilibrium}, computation of equilibria is discussed.
In Section~\ref{sec:linearization}, the nonlinear model is linearized around different operating points.
In Section~\ref{sec:controller_synthesis}, the linearized models are "shaped" and $\hinf$ controllers for the different operating points are synthesized.
Finally, in Section~\ref{sec:reference_generation}, the control loop is summarized.

\subsection{Equilibria}
\label{sec:equilibrium}
Equilibria of the dynamics (\ref{eq:state_equation}) are characterized by a steady state ${x^o=\begin{bmatrix}\omega^o&x_t^o&v_t^o\end{bmatrix}^T}$, where $\dot{x}(t) = 0$, and a corresponding output $y^o=\begin{bmatrix}\omega^o& \lambda^o& P^o& x_t^o\end{bmatrix}^T$ for fixed values of the control input, $u^o=\begin{bmatrix}\theta^o&M_g^o\end{bmatrix}^T$, and uncertain input, $V(t)=V^o$.
For power maximization, the steady state is exclusively determined by the wind speed, whereas for power reference tracking, the steady state depends on both wind speed and power reference $P^\text{ref}(t)\in\Rp$ as will be discussed in what follows.

\subsubsection{Power Maximization}
The maximum power is achieved for a maximum $C_p(\cdot)$.
Therefore, we choose $(\lambda^o,\theta^o) = \argmax_{\lambda, \theta} C_P(\lambda, \theta)$.
Due to the particular shape of $C_P(\lambda,\theta)$, a global maximum $(\lambda^o, \theta^o)$ uniquely exists independent of the wind speed \cite{Hansen2008}.
This allows us to deduce $\omega^o=\frac{N_g\lambda^o V^o}{r}$.
The associated generator torque $M_g^o$ can then be computed from (\ref{eq:state_equation}) via $\dot{\omega}(t) = 0$ as
\begin{equation}
	M_g^o = \frac{\rho\pi r^2}{2}\frac{(V^o)^3}{\omega^o}C_P(\lambda^o, \theta^o),
	\label{eq:generator_speed_equilibrium}
\end{equation}
and the equilibrium power as $P^o = \eta\omega^oM_g^o$.
The steady state conditions $\dot{v}_t(t) = 0$, $\dot{x}_t(t) = v_t^o = 0$, yield
\begin{equation}
	x_t^o = \frac{\rho\pi r^2}{2K_t}(V^o)^2C_T(\lambda^o, \theta^o).
	\label{eq:equilibrium_tower_position}
\end{equation}

\subsubsection{Power Reference Tracking}
For power tracking, we use an equilibrium that matches the reference $P^\text{ref}$, i.e.,
\begin{equation}
	P^\text{ref} = P^o = \eta\omega^o M_g^o.
	\label{eq:equilibrium_power}
\end{equation}
From the equilibrium condition ${\dot{\omega}(t)=0}$, we deduce
\begin{equation}
	M_g^o = \frac{\rho\pi r^2}{2}\frac{(V^o)^3}{\omega^o}C_P\Big(\frac{r}{N_g}\frac{\omega^o}{V^o}, \theta^o\Big),
	\label{eq:steady_state_generator}
\end{equation}
from (\ref{eq:state_equation}).
Combining (\ref{eq:equilibrium_power}) and (\ref{eq:steady_state_generator}) results in
\begin{equation}
	C_P\Big(\frac{r}{N_g}\frac{\omega^o}{V^o}, \theta^o\Big) = \frac{2P^o}{\rho\pi r^2\eta(V^o)^3}.
	\label{eq:omega_theta_condition}
\end{equation}
For a given value of the right hand side of (\ref{eq:omega_theta_condition}), we find a solution $\theta^o$, by using a lookup table that best relates output power and generator speed \cite{jeong2014comparison}, i.e.,
\begin{equation}
	\omega^o = \operatorname{LUT}(P^o).
\end{equation}
Evidently, (\ref{eq:equilibrium_tower_position}) also holds for this case with $\lambda^o = \frac{r}{N_g}\frac{\omega^o}{V^o}$.

\subsection{Linearization}
\label{sec:linearization}
The controllers for the nontrivial operating regions $i\in\{2,3\}$ rely on the same nonlinear model (\ref{eq:nonlinear_model}).
We refer equilibria in region $i$ by $(x_i^o, u_i^o, V_i^o)$.
Let $\xi_i(t) = x(t) - x_i^o$, $\mu_i(t) = u(t) - u_i^o$, and $\nu_i(t) = y(t) - y_i^o$ denote the deviations from the respective equilibria.
The linear models
\begin{subequations}
\begin{equation}
	\dot\xi_i(t)=A_i\xi_i(t)+B_i\mu_i(t),
\end{equation}
\begin{equation}
	\nu_i(t)=C_i\xi_i(t)+D_i\mu_i(t),
\end{equation}
\end{subequations}
can then be derived by computing the Jacobians
\begin{subequations}
\begin{equation}
	A_i =\left.\textstyle\frac{\partial f\left(\ldots\right)}{\partial x(t)}\right|_{(x_i^o, u_i^o, V_i^o)},\quad
	B_i =\left.\textstyle\frac{\partial f\left(\ldots\right)}{\partial u(t)}\right|_{(x_i^o, u_i^o, V_i^o)},\\
\end{equation}
\begin{equation}
	C_i =\left.\textstyle\frac{\partial g\left(\ldots\right)}{\partial x(t)}\right|_{(x_i^o, u_i^o, V_i^o)},\quad
	D_i =\left.\textstyle\frac{\partial g\left(\ldots\right)}{\partial u(t)}\right|_{(x_i^o, u_i^o, V_i^o)}.
\end{equation}
\label{eq:linear_model}
\end{subequations}
The associated nominal \acp{tfm} are
\begin{equation}
    G_i^n(s) = C_i(s I_3 - A_i)^{-1}B_i + D_i.
\end{equation}

\subsection{Controller Synthesis}
\label{sec:controller_synthesis}
To synthesize controller $K_i(s)$ for $G_i^n(s)$, we use the $\hinf$ loop-shaping approach described in \cite{mcfarlane1990robust} (see also \cite{raisch1994mehrgrossenregelung,skogestad2005multivariable}).
Given a nominal model $G_i^n(s)$, we use asymptotically stable pre- and post-compensator \acp{tfm} $W_i^\text{pre}(s)$ and $W_i^\text{post}(s)$, resulting in the augmented plant
\begin{equation}
    G_i^a(s) = W_i^\text{post}(s)G_i^n(s)W_i^\text{pre}(s).
\end{equation}
Pre- and post-compensators are chosen to reflect design specifications.
In particular, the singular values of $G_i^a(j\omega)$, ${\omega\geq 0}$, can be considered "targets" for the open loop frequency response singular values, i.e., the singular values of $G_i^n(j\omega)K_i(j\omega)$.
According to \cite{mcfarlane1990robust}, synthesis of the controller \ac{tfm} proceeds as follows:
\begin{enumerate}[(a)]
	\item Determine a normalized left coprime factorization of the augmented \ac{tfm} $G_i^a(s)$, i.e., asymptotically stable \acp{tfm} $M_i^a(s)$, $N_i^a(s)$ s.t. $G_i^a(s) = M_i^a(s)^{-1}N_i^a(s)$, there are no right half plane pole-zero-cancellations when forming ${M_i^a}^{-1}N_i^a$, and $M_i^a(s)M_i^a(-s)^T + N_i^a(s)N_i^a(-s)^T = I$. 
	From a minimal realization of $G_i^a(s)$, it is straightforward to compute (a minimal realization of) $\begin{bmatrix}N_i^a(s)& M_i^a(s)\end{bmatrix}$ (see \cite{mcfarlane1990robust}).
	\item Consider the $\hinf$ minimization problem
	\begin{equation}
    	\min_{K_i^a(s)}\underbrace{\left\|\begin{bmatrix}K_i^a(s)\left(I-G_i^a(s)K_i^a(s)\right)^{-1}{M_i^a(s)}^{-1}\\\left(I-G_i^a(s)K_i^a(s)\right)^{-1}{M_i^a(s)}^{-1}\end{bmatrix}\right\|_\infty}_{\gamma_i}\!\!\!\!
	\end{equation}
	where minimization is over all realizable \ac{tfm} $K_i^a(s)$ that stabilize the shaped nominal plant model $G_i^a(s)$.
	The minimal cost $\gamma_{i,\text{min}}$ can be computed analytically \cite{mcfarlane1990robust}.
	As shown in \cite{mcfarlane1990robust}, large values of $\gamma_{i,\text{min}}$, i.e., $\gamma_{i,\text{min}}\gg 1,$ are an indicator that the design specification expressed via $W_i^\text{post}(s)$ and $W_i^\text{pre}(s)$ are not achievable and suggest to revise the choice of pre- and post-compensator.
	\item To avoid numerical subtleties, one can compute a (slightly) suboptimal \ac{tfm} $K_i^a(s)$, achieving a cost
	\begin{equation}
		\gamma_{i,\text{sub}} > \gamma_{i,\text{min}}, \quad\frac{\gamma_{i,\text{sub}} - \gamma_{i,\text{min}}}{\gamma_{i,\text{min}}} \ll 1.
	\end{equation}
	An algorithm to analytically compute (a state model realization of) $K_i^a(s)$ is also given in \cite{mcfarlane1990robust}.
	\item From the controller for the augmented plant model $K_i^a(s)$, one recovers $K_i(s)$, the controller for the "original" plant model $G_i^n(s)$ via
	\begin{equation}
		K_i(s) = W_i^\text{pre}(s)K_i^a(s)W_i^\text{post}(s).
	\end{equation}
	\item Controllers $K_i^a(s)$, respectively $K_i(s)$, by construction, stabilize the nominal plant model $G_i^a(s)$, respectively $G_i^n(s)$.
	Moreover, uncertain plant dynamics of the form
	\begin{equation}
		G_i^{p,s}(s) = \left(M_i^a(s) + \Delta M_i(s)\right)^{-1}\left(N_i^a(s) + \Delta N_i(s)\right),
	\end{equation}
	where $\Delta M_i(s)$, $\Delta N_i(s)$ are arbitrary asymptotically stable unknown \acp{tfm} with
	\begin{equation}
		\left\|\begin{bmatrix}\Delta N_i &\Delta M_i\end{bmatrix}\right\|_\infty\leq\frac{1}{\gamma_{i,\text{sub}}},
	\end{equation}
    do not affect closed loop stability \cite{mcfarlane1990robust}.
\end{enumerate}

\subsection{Linear Controller for Nonlinear Plant}
\label{sec:reference_generation}
\begin{figure}[htbp]
	\centering
	\includegraphics{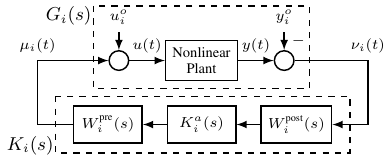}
	\caption{Loop-shaping design approach: Closed-loop $\hinf$-control diagram.}
	\label{fig:closed_loop_nonlinear}
\end{figure}

As discussed in Section \ref{sec:equilibrium}, suitable equilibria are computed for operating region $i$, $i\in\{2,3\}$, depending on the current wind speed and, for $i=3$, the power reference value.
As shown in Figure~\ref{fig:closed_loop_nonlinear}, the output of the linear controller $K_i(s)$ is added to the equilibrium control input $u_i^o$ to give the control signal $u(t)$, while the input of the controller $K_i(s)$ is the deviation of the plant output $y(t)$ from the equilibrium output $y_i^o$.


\section{Bumpless Transfer}
\label{sec:transfer}
In Section~\ref{sec:controller}, we designed separate controllers for two operating regions, resulting in power maximization controller $K_2$ and power reference tracking controller $K_3$.
Under realistic operating conditions, the wind speed may fluctuate such that transitioning between power maximization and power tracking becomes necessary.
Here, we will discuss this transition.

\begin{figure}[b]
	\centering
    \includegraphics{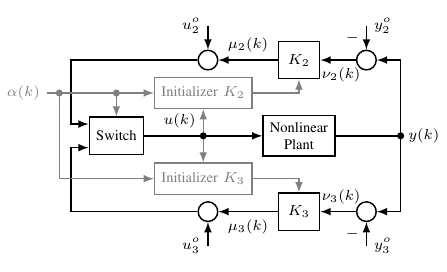}
	\caption{Bumpless switching with $K_2$ and $K_3$ running in parallel.}
	\label{fig:bumpless_switching}
\end{figure}

For the remainder of this work we will use discrete-time notation with time instance $k\in\N_0$ to bridge the gap to applications on digital controllers.
In Figure~\ref{fig:bumpless_switching}, the closed control loop is shown where controllers $K_2$ and $K_3$ run in parallel and a switch decides, based on the control mode $\alpha(k)\in\{2,3\}$, which control signal to use.
The controller is selected based on $V(k)$ and $P^\text{ref}(k)$, with margin $\beta_\text{rel}$, i.e.,
\begin{equation}
    \alpha(k) = \begin{cases}3, &\text{if }\frac{\rho\pi r^2\eta}{2}V(k)^3C_P(\lambda^o, \theta^o) > \beta_\text{rel}P^\text{ref}(k),\\
    2, &\text{else.}
    \end{cases}
\end{equation}
Hence, if the available wind power exceeds the power reference with margin, $K_3$ is applied.
Otherwise, $K_2$ is used to maximize the power production.
Note that, in practice for power tracking to work a margin $\beta_\text{rel} > 1$ is required.

We employ the bumpless transfer scheme described in \cite{paxman2000optimal} to ensure a smooth transition from $K_i$ to $K_j$, $i,j\in\{2,3\}$, ${i\neq j}$.
Consider dynamics of controller $i$ of the form
\begin{subequations}
\begin{align}
    \xi_{c,i}(k+1) = \hat{A}_i\xi_{c,i}(k) + \hat{B}_i\nu_i(k),\\
    \mu_i(k) = \hat{C}_i\xi_{c,i}(k) + \hat{D}_i\nu_i(k).
\end{align}
\end{subequations}
At switching time $\kappa\in\N$, we set the initial state $\xi_{c,j}(\kappa)$ for controller $K_j$ such that hypothetical past input $\mu_j(\kappa-1) + u_j^o$ and output $\nu_j(\kappa-1) + y_j^o$ match the actual past input and output $u(\kappa-1)$ and $y(\kappa-1)$ as closely as possible.
If we define close in the sense of the Euclidean vector norm, the desired initial state $\xi_{c,j}(\kappa)$ is found by solving the following optimization problem.
\begin{problem}
    \label{pro:optimization_problem}
    \begin{multline*}
        \min_{\substack{\xi_{c,j}(\kappa),\xi_{c,j}(\kappa-1),\\\nu_j(\kappa-1),\mu_j(\kappa-1)}}\textstyle\left\|\begin{bmatrix}\mu_j(\kappa-1)+\mu_j^o-u(\kappa-1)\\\nu_j(\kappa-1)+\nu_j^o-y(\kappa-1)\end{bmatrix}\right\|_2^2
    \end{multline*}
    subject to
    \begin{subequations}
    \begin{align}
        \xi_{c,j}(\kappa) &= \hat{A}_j\xi_{c,j}(\kappa-1) + \hat{B}_j\nu_j(\kappa-1),\\
        \mu_j(\kappa-1) &= \hat{C}_j\xi_{c,j}(\kappa-1) + \hat{D}_j\nu_j(\kappa-1).
    \end{align}
    \end{subequations}
\end{problem}
Here, the variables $\mu_j^o$, $\nu_j^o$, $u(\kappa-1)$, and $y(\kappa-1)$ are known.
The optimal controller state $\xi_{c,j}^\ast(\kappa)$ will then be used as an initial value for controller $j$ at switching instance $\kappa$.
Thereby, smooth transient behavior is ensured for the combined control scheme.
Note, Problem \ref{pro:optimization_problem} can be cast as a least squares problem which can be solved analytically \cite{adaptive1991astrom}.


\section{Case Study}
\label{sec:case_study}
We investigate the control strategy described above by applying it to the OpenFAST model of the IEA \SI{3.4}{\mega\watt} reference wind turbine in different scenarios \cite{RWT}.
For comparison, the reference open source controller (ROSCO) is used with the default parameters \cite{abbas2022reference}.
Note that, while original ROSCO is not capable of tracking a power reference signal, we adapt it by adjusting the rated power and rated generator speed dynamically depending on the power reference \cite{kim2018design}.

In Section~\ref{sec:cs_simulation_setup}, the simulation setup is discussed.
In Section~\ref{sec:cs_robustness} robustness is analyzed.
In Section~\ref{sec:cs_powertracking}, power maximization and power tracking performance are investigated.
In Section~\ref{sec:cs_switching}, the bumpless transfer is evaluated.
And Section~\ref{sec:cs_del} discusses \acp{del}.

\subsection{Simulation Setup}
\label{sec:cs_simulation_setup}
Analogous to prior work \cite{grapentin2024robust,grapentin2022lq}, we use the nonlinear {OpenFAST} model with saturation and slew rate constraints on the inputs to evaluate the controllers.
All simulations are executed with sampling time $T_s = \SI{4}{\milli\second}$.
The controller relies on wind speed estimates from the wind speed observer presented in \cite{schreiber2020field}.

For the synthesis of controllers $K_2$ and $K_3$, the following pre- and post-compensator are chosen
\begin{align*}
    W_2^\text{pre}&=\diag\left(\textstyle\frac{5.2}{s+2}, \frac{1579}{s+50}\right),
    W_3^\text{pre}=\diag\left(\textstyle\frac{10.4}{s+2}, \textstyle\frac{6.315}{s+2}\right),\\
    W_2^\text{post} &= \diag\left(7.6\cdot 10^{-5}, \textstyle\frac{0.5s+0.25}{0.01s^2+s}, \frac{2.9\cdot 10^{-12}}{10^2s+1}, \frac{0.01}{10s+1}\right),\\
    W_3^\text{post} &= \diag\left(\textstyle \frac{6.1s+0.76}{10^3 s}, 5\cdot 10^{-11}, \frac{1.18s+2.37}{2\cdot 10^5s}, \frac{10^{-4}}{100s+1}\right).
\end{align*}
The pre-compensators include the first order actuator dynamics.
Whereas $W_i^\text{post}$ are tailored to the individual control objectives:
For power maximization, $W_2^\text{post}$ is setup to primarily reduce the tip-speed ratio error using a PI-element.
For power reference tracking, $W_3^\text{post}$ has PI-elements for the generator speed and the output power, resulting in offset free power tracking while staying below the upper generator speed bound.
The weights for all other outputs are chosen with low gain.

\subsection{Robustness}
\label{sec:cs_robustness}
When synthesizing controllers $K_i$, $i\in\{2,3\}$, the resulting cost $\gamma_{i,\text{sub}}$ is the inverse of the coprime uncertainty level for which stability can be guaranteed (see Section \ref{sec:controller_synthesis}).
In our case study, we have $\nicefrac{1}{\gamma_{2,\text{sub}}} = 0.61$ and $\nicefrac{1}{\gamma_{3,\text{sub}}} = 0.64$.
Linearizing the wind turbine model (\ref{eq:state_equation}) around different equilibria in both regions and shaping all models in each region with the same pre- and post-compensators results in perturbed systems $G_{i,j}^{p,s}(s)$, $i\in\{2,3\}$, $j\in\{1,2,\dots\}$.
From the normalized left coprime factorization, we can then compute the error \acp{tfm} $\Delta M_{i,j}(s)$ and $\Delta N_{i,j}(s)$, and assess if the robustness condition
\begin{equation}
    \left\|\begin{bmatrix}\Delta N_{i,j}& \Delta M_{i,j}\end{bmatrix}\right\|_\infty\leq\frac{1}{\gamma_{i,\text{sub}}},\label{eq:robustness_condition}
\end{equation}
holds.
If this is the case, controller $K_i$ will stabilize the plant models obtained from linearization around the equilibrium points $j\in\{1,2,\dots\}$.

\begin{figure}[htbp]
	\centering
	\includegraphics{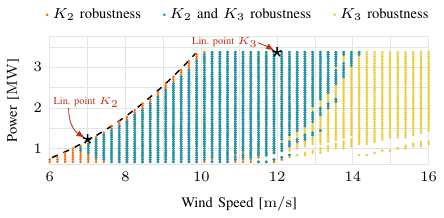}
	\caption{Robustness analysis for $K_2$ and $K_3$ indicate an overlapping area (in blue) where both controllers guarantee stability. Note that all operating points lie below the dashed line that represents the physical maximum power that can be harvested at a given wind speed.}
	\label{fig:robustness_analysis}
\end{figure}

In Figure~\ref{fig:robustness_analysis}, different operating points are sampled, resulting in different $\Delta M_{i,j}$ and $\Delta N_{i,j}$, and (\ref{eq:robustness_condition}) is evaluated for controllers $K_2$ and $K_3$, respectively.
Both controllers satisfy (\ref{eq:robustness_condition}) for large areas of operating points.
As the power maximization controller $K_2$ operates on or slightly below the dashed line, robustness is ensured at every wind speed when maximizing the output power.
For controller $K_3$, robustness is ensured in a large area around its linearization point.
Moreover, there is a large area (shown in blue) where both controllers ensure robust stability and bumpless transfer can be performed.
Note that robustness can be varied by appropriately choosing the linearization points and the pre-/post-compensators for the synthesis of $K_2$ and $K_3$.

\subsection{Power Maximization and Power Reference Tracking}
\label{sec:cs_powertracking}
The primary aim of the presented controller is to achieve highly accurate power reference tracking while being competitive at power maximization.
\begin{figure}[b]
	\centering
	\includegraphics{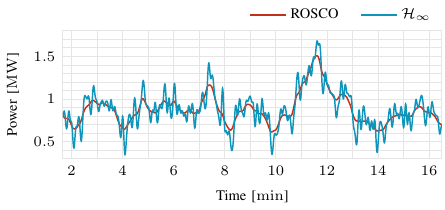}
	\caption{Power maximization below rated wind speed at a mean wind speed of \SI{6.4}{\meter\per\second} and a turbulence intensity of \SI{7}{\percent}.}
	\label{fig:powerTrackingLow}
\end{figure}
We compare the presented controller with a state-of-the-art ROSCO controller \cite{abbas2022reference}.
In Figure~\ref{fig:powerTrackingLow}, the resulting output power for a low wind speed scenario is shown.
Note that no switching occurs and controller $K_2$ is continuously active in the $\hinf$ design.
The trajectories are very similar, with larger variations for the $\hinf$ controller.
While the ROSCO controller produces a smoother power output signal, the $\hinf$ controller harvests slightly more energy in total.
\begin{figure}[t]
	\centering
	\includegraphics{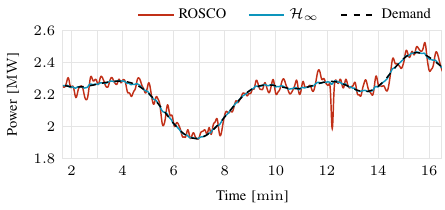}
	\caption{Power tracking above rated wind speed at a mean wind speed of \SI{14.8}{\meter\per\second} and a turbulence intensity of \SI{6}{\percent}.}
	\label{fig:powerTrackingHigh}
\end{figure}
Figure~\ref{fig:powerTrackingHigh} displays a high wind speed scenario, with the ROSCO and the $\hinf$ controller tracking a power reference.
While the ROSCO has a root mean square tracking error of \SI{29.21}{\kilo\watt}, the $\hinf$ controller achives a tracking error of \SI{3.46}{\kilo\watt}.
This shows an 8-times improvement of the tracking accuracy.

\subsection{Bumpless Transfer}
\label{sec:cs_switching}
\begin{figure}[htbp]
	\centering
	\includegraphics{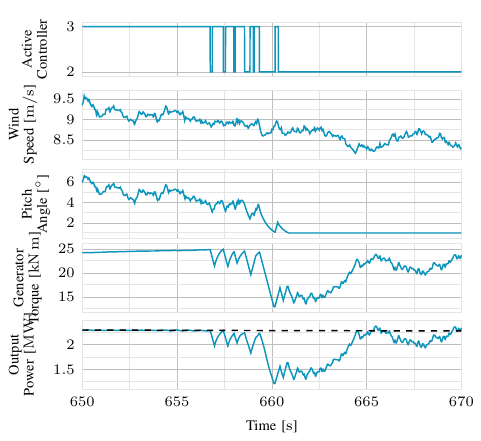}
	\caption{$\hinf$ Controller switching while transitioning from power reference tracking to power maximization. The blue lines denote measurements and the dashed line denotes the power reference. Controller mode 2 denotes power maximization while mode 3 corresponds to power reference tracking.}
	\label{fig:controllerTransfer}
\end{figure}

We now simulate a scenario where wind speed changes significantly, requiring switching between power maximization and power tracking.
Figure~\ref{fig:controllerTransfer} displays the active controller, wind speed, pitch angle, generator torque, and output power.
In the first \SI{655}{\second} of the simulation, the wind speed was high and power reference tracking was possible.
After \SI{655}{\second}, there is a period, where a number of changes between $K_3$ and $K_2$ is required.
From \SI{665}{\second}, the wind has slowed down such that the power maximization controller $K_2$ is active continuously.
Despite the oscillation from \SI{655}{\second} to \SI{665}{\second}, there are no large jumps in the pitch angle and generator torque because of the bumpless transfer scheme that ensures smooth transitions.
Moreover, the power overshoot remains small given the strongly fluctuating operating conditions.

\subsection{Damange Equivalent Loads}
\label{sec:cs_del}
\begin{figure}[htbp]
	\centering
	\includegraphics{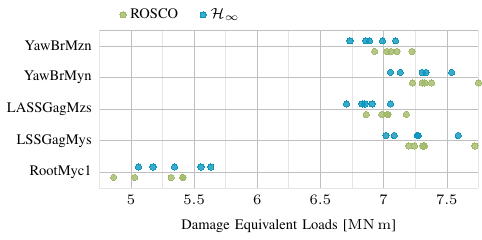}
	\caption{Damage equivalent loads for ROSCO and switching $\hinf$ controllers. The blade root out-of-plane bending moment (RootMyc1), shaft non-rotating out-of-plane bending moment (LSSGagMys), shaft non-rotating yaw bending moment (LSSGagMzs), tower top fore-aft bending moment (YawBrMyn), and tower top torsion bending moment (YawBrMzn) are displayed.}
	\label{fig:del_boxplot}
\end{figure}

\acp{del} are used to quantify mechanical stress by extrapolating the load cycles onto the wind turbine lifetime.
They allow us to compare the mechanical wear associated with different controllers indicating if maintenance is required more frequently.
In numerical simulations, \acp{del} depend on the specific seed used to generate wind \cite{jonkman2022openfast}.
Therefore, we ran each simulation using five different seeds, leading to different \acp{del}.
In Figure~\ref{fig:del_boxplot} the \acp{del} are shown for a selection of components.
The ROSCO and the switching $\hinf$ controllers achieve very similar mechanical loads, with the $\hinf$ controller achieving slightly better results for tower.


\section{Conclusion}
\label{sec:conclusion}
We have presented a combined power maximization/power tracking control scheme based on $\hinf$ loop-shaping with bumpless controller transfer between both operating modes.
We provide a robustness criterion, which allows us to guarantee stability in a large number of operating points.
The control scheme achieves highly accurate power reference tracking with slightly decreased mechanical tower loads in comparison with the ROSCO controller.


\addtolength{\textheight}{-0cm}   




\bibliography{literature}

\end{document}